\documentclass[preprint]{ccn}  % options: submission (default), preprint, proceedings, extended-abstract

\usepackage{stfloats}
\usepackage{float}

\title{Conserved Kinematic Representations enable Zero-Shot Decoding in Handwriting BCIs}

\author{%
  Srinivas Ravishankar\affmark{1} \And
  Virginia de Sa\affmark{1, 2}
  }
\affiliation{1}{Department of Cognitive Science, UC San Diego}
\affiliation{2}{Halicioglu Data Science Institute, UC San Diego}
\emails{srravishankar@ucsd.edu}

\begin{document}

\maketitle

\begin{abstract}
While intracortical Brain-Computer Interfaces (iBCIs) that decode imagined handwriting have achieved high communication rates for Latin scripts, they rely on observing every character in the alphabet during training. This poses a challenge in scaling to logographic languages (e.g., Chinese, Japanese), where the character set exceeds thousands of classes. The limitation highlights a fundamental question in motor neuroscience: does the motor cortex represent handwriting through the composition of shared kinematic primitives, that can be exploited by decoders? We introduce a computational framework for aligning neural activity to imagined kinematics in large datasets, enabling the training of a zero-shot capable machine learning algorithm for decoding unseen characters. Our model achieves 64\% hits@3 retrieval on unseen letters, suggesting that neural representations of kinematic strokes are robustly conserved across different character contexts. 
% We further analyze the neural sources driving this generalization, isolating a low-dimensional subspace of "stable" information that encodes motor primitives invariant to sequence identity. 
This study provides a framework for dissecting conserved neural dynamics in large-scale intracortical datasets and offers strong evidence for a compositional basis of complex motor control. It also establishes a new paradigm for open-vocabulary iBCI communication with minimal recalibration burden on the user, crucial to increasing adoption of neuroprosthetics in logographic languages.
\end{abstract}

\section{Introduction}
\label{sec:introduction}
Individuals with severe paralysis or speech disorders face significant challenges in communication. 
Brain-computer interfaces (BCIs) translate brain signals directly into text or speech, bypassing traditional peripheral systems such as the hands or voice. Handwriting BCIs in particular decode neural activity related to imagined writing (mentally tracing penstrokes without overt movement). Such a system was demonstrated by \citet{willett2021high} to achieve state-of-the-art accuracy and allow intuitive, high-throughput communication for those unable to use traditional  modes of communication. Furthermore, these technologies are poised to offer new modes of interaction for healthy users as well. 

However, we anticipate key challenges in their adoption. Existing methods employ relatively simple machine learning methods to map neural activity to recognized characters, where neural training data must be collected for each character. Although they have been demonstrated in English with a limited character set, it is non-trivial to scale to large-character languages such as Chinese or  Japanese, which requires $\sim 2500$ characters for regular use. This requirement blocks access to these technologies for large segments of the population.

\begin{figure}[h]
    \centering
    \includegraphics[width=0.4\textwidth]{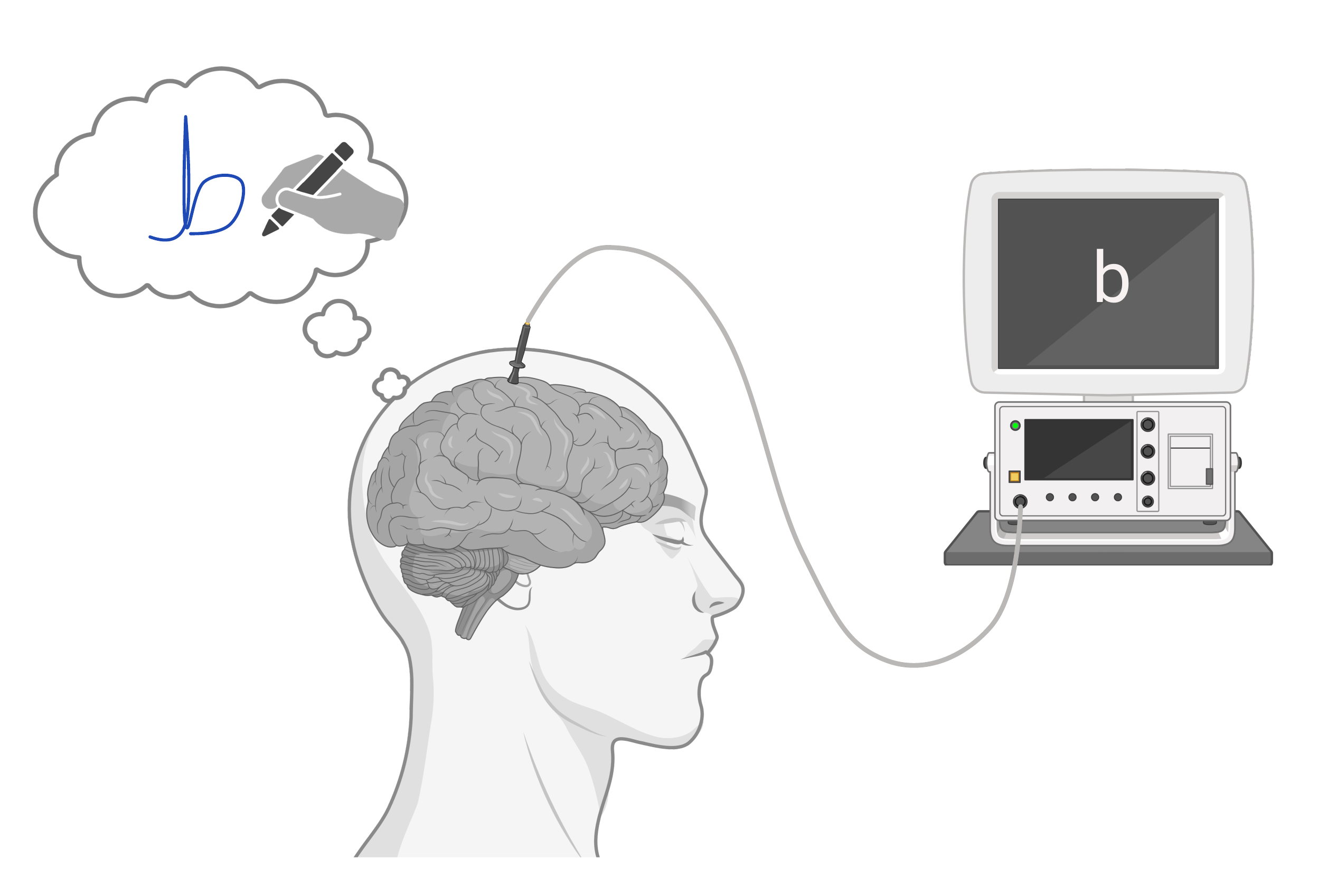}
    \caption{High performance has been demonstrated when decoding imagined handwriting in English, with a limited character set. Logographic languages may require more sophisticated methods.}
    \label{fig:imagined_handwriting}
\end{figure}%

To address these limitations, we propose to enable zero-shot character decoding in handwriting BCIs to reduce data demands in existing systems, and extend usability to language with large character sets. 

While benchmarks such as FALCON \citep{karpowicz2024few} have been proposed to evaluate few-shot and zero-shot generalization to new sessions, a different dimension of zero-shot generalization has received very limited attention: decoding of new characters unseen during training.

Some recent works have explored zero-shot handwriting decoding in Chinese \citep{ yang2024reconstructing, qi2025human}, but are limited to slow single-letter writing. The participant was asked to imagine writing with their entire arm, and each character took 4 to 9 seconds to write. No existing system supports ballistic continuous handwriting-to-text comparable to the performance previously demonstrated by \citet{willett2021high} while allowing zero-shot capability.

Furthermore, prior work requires supervised single-letter data collection for initial training as well as recalibration. Recalibration is a continuous burden on BCI users, arising due to non-stationarity and electrode shift. Neural signals are inherently non-stationary; their statistical distribution changes across sessions due to biological noise or plasticity, impedance fluctuations, and physical electrode shifts relative to the brain tissue. These phenomena render the decision boundaries of the initial classifier increasingly obsolete with time, and necessitate the collection of new labeled data to restore accuracy. It is crucial to minimize recalibration requirements to have a functional BCI for daily use by participants. 

In this study, we demonstrate a method for handwriting decoding that is zero-shot capable, and can be trained and recalibrated without any supervised single-letter data collection. This is achieved through kinematics prediction,
and building on unsupervised recalibration work introduced for handwriting decoding \citep{fan2023plug}. 

% As proof of concept of its efficacy, the method is evaluated by cutting out all snippets of \textcolor{red}{ALL} character in the training data of the existing handwriting dataset.

While an imagined logographic dataset is not yet publicly available, we demonstrate a proof of concept in this work. We evaluate this method by cutting out all snippets of each character in the training data of an existing English imagined handwriting dataset to simulate the zero-shot setting.

\section{Methods}
\label{sec:methods}

\subsection{Dataset}

The neural dataset used in this study was obtained from intra-cortical micro-electrode recordings during imagined handwriting by a single participant in a prior study \citep{willett2021high}.
Data were collected using 2 Utah arrays implanted in the hand knob area of the precentral gyrus, with a total of 192 electrodes over 10 sessions and preprocessed in the same manner as the original work: Multi-Unit threshold crossing rates were computed, and binned into 20ms time steps. Data were smoothed with a causal Gaussian filter. Each block of data was mean-subtracted and divided by the standard deviation per electrode. 

We follow the same training and validation trial split as the original work to train continuous sentence classification models. For our zero-shot character recognition experiments, we use all data across training and validation trials, but ensure that all neural snippets associated with the chosen zero-shot character is excluded. Zero-shot character recognition performance is evaluated on the held-out character snippets from both training and validation trials.

% This study also uses online handwriting data can be collected from hundreds of healthy participants in order to recognize unseen characters. Handwriting recognition data is obtained from the Deepwriting dataset \cite{aksan2018deepwriting}.
% % , an extended version of the popular IAM-OnDB \cite{liwicki2005iam} datatset. 
% The dataset contains a total of 85,560 word instances across 294 unique authors, with character-level annotations (406, 956 characters). 
% % represented by sequences of tuples containing x-coordinate, y-coordinate and pen-up events. 
% We modify this dataset to match the statistics of the neural dataset, replacing spaces and periods with synthetic trajectories representing `$>$' and `$\sim$' trajectories 
% %instead of 
% respectively to improve transfer learning.

% We demonstrate an offline proof of concept by holding out data from the  

\subsection{Architecture}

\begin{figure*}[ht]
	\centering
    \includegraphics[width=0.9\textwidth]{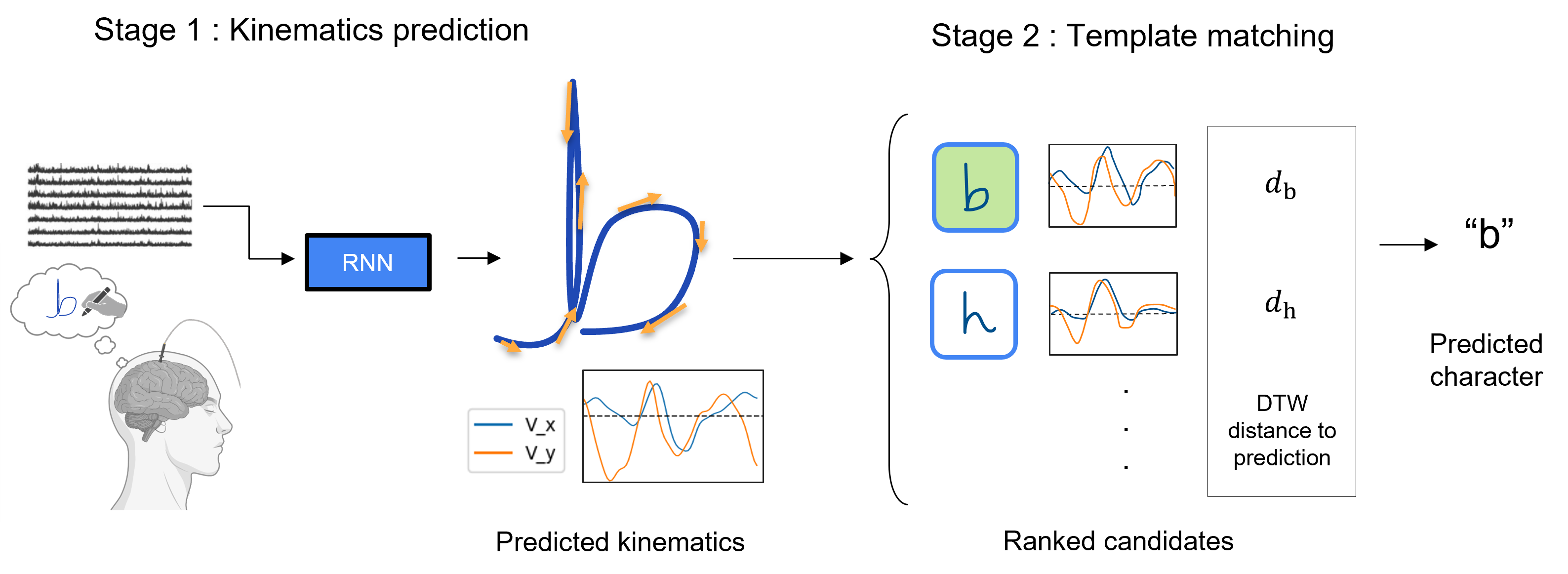}
    \caption{Two-stage pipeline to enable zero-shot character recognition. The first stage (kinematics prediction) predicts a velocity sequence of a hypothetical pen-tip corresponding to the observed neural data. The second stage (template matching) ranks a character template dictionary using a DTW-based distance measure, enabling zero-shot recognition.}
    \label{fig:two_stage_pipeline}
\end{figure*}

To create a zero-shot capable system, we propose a two-stage architecture shown in Fig \ref{fig:two_stage_pipeline}. The first stage maps neural data to a velocity sequence of a hypothetical pen-tip. While we use a Recurrent Neural Network (RNN) for this stage, the architecture is agnostic to the specific choice of sequence model. The second stage uses a template matching approach to find the most likely candidate, given a library of kinematics sequences for different characters. 

% \subsubsection{Kinematics Prediction}

\vspace{3mm}

There are significant challenges that need to be solved to train the kinematics prediction model. Given continuous sentence data, we lack the full ground truth kinematics of the entire sentence (consider the velocity profile for the gaps between characters). Moreover, even for characters where we have approximate ground truth of the participant's imagined writing, we lack precise frame-wise timing. 

Due to this variability in timing between the predicted sequence $\mathbf{\hat{y}}$ and the approximate ground truth $\mathbf{y}$, we  employ a soft Dynamic Time Warping loss~\citep{cuturi2017soft} to optimize our kinematics model. 

% $$DTW(X, Y) = \min_{W} \left[ \sum_{k=1}^{K} \delta(w_k) \right]$$

Unlike element-wise metrics such as Mean Squared Error, DTW aligns sequences by finding the optimal non-linear warping path that minimizes the cumulative alignment cost. However, the classical DTW formulation relies on a discrete $\min$ operator, which is non-differentiable and unsuitable for gradient-based optimization.

Soft-DTW addresses this by replacing the hard minimum with a generalized $\text{softmin}_\gamma$ operator, defined as:
\begin{equation}
    \text{softmin}_\gamma(z_1, \dots, z_k) = -\gamma \log \sum_{i=1}^k \exp\left(-\frac{z_i}{\gamma}\right),
    \label{eq:softmin}
\end{equation}
where $\gamma > 0$ is a smoothing hyperparameter. As $\gamma \to 0$, the objective converges to the exact DTW solution. The loss is computed via dynamic programming; let $\Delta_{i,j} = \| \hat{y}_i - y_j \|^2$ represent the local cost between time steps $i$ and $j$. The cumulative soft cost matrix $R$ is derived recursively:
\begin{equation}
    R_{i,j} = \Delta_{i,j} + \text{softmin}_\gamma(R_{i-1, j}, R_{i, j-1}, R_{i-1, j-1}).
    \label{eq:softdtw_recursion}
\end{equation}

After the entire cumulative cost matrix R is computed, the final Soft-DTW loss is simply the value in the bottom-right cell of the matrix
$R_{n,m}$, where n is the length of sequence $\mathbf{\hat{y}}$ and m is the length of sequence $\mathbf{y}$.
The Soft-DTW loss computed above is fully differentiable with respect to the inputs, enabling end-to-end training via standard backpropagation.

% We first extract character-level snippets from the neural data, using a simple method which is novel to the best of our knowledge to automatically recalibrate the kinematics prediction model.

However, note that the dynamic programming solution scales quadratically with sequence length. Thus, naively optimizing long continuous sentence data with partial ground truth poses challenges due to both computational tractability and poor learning signals from partial alignment. The solution is to extract character-level snippets from the continuous sentence data, to reduce computational complexity and to make the learning problem well-behaved.

Here we describe a simple novel method to extract such snippets from continuous sentence data, allowing kinematics training. We use models already employed for continuous character recognition in the handwriting decoding task, described below:  

Prior work \citep{fan2023plug} used recurrent models trained using Connectionist Temporal Classification (CTC) to decode handwriting from neural data.
More generally, CTC is used to map input features $X$ to label sequences $Y$ without requiring explicit frame-level alignment. CTC extends the vocabulary $V$ with a `blank' token $\epsilon$, predicting a probability distribution over $V \cup \{\epsilon\}$ at each time step.

A many-to-one mapping function $\mathcal{B}$ recovers the final sequence by merging repeated characters and removing blanks. The network is trained to minimize the negative log-likelihood of the ground truth $Y$ by marginalizing over all valid alignment paths $\pi$:

\begin{equation}
    \mathcal{L}_{CTC} = - \ln \sum_{\pi \in \mathcal{B}^{-1}(Y)} \prod_{t=1}^{T} p(\pi_t \mid t, X)
\end{equation}

This summation is computed efficiently via the forward-backward algorithm, enabling end-to-end differentiability.

However, CTC-trained models cannot traditionally be used for segmentation due to so-called `peaky' behavior, which has been previously discussed in \citet{zeyer2021does}. The model produces a sharp probability peak when a character is recognized, but does not explicitly denote the duration of any character. A causal model produces a peak at the end of each recognized label, when it accumulates sufficient evidence to distinguish it from other characters. A non-causal model, such as a bidirectional RNN, will instead produce such a peak at some arbitrary time within the span of a class.

\begin{figure*}[t]
	\centering
    \includegraphics[width=0.75\textwidth]{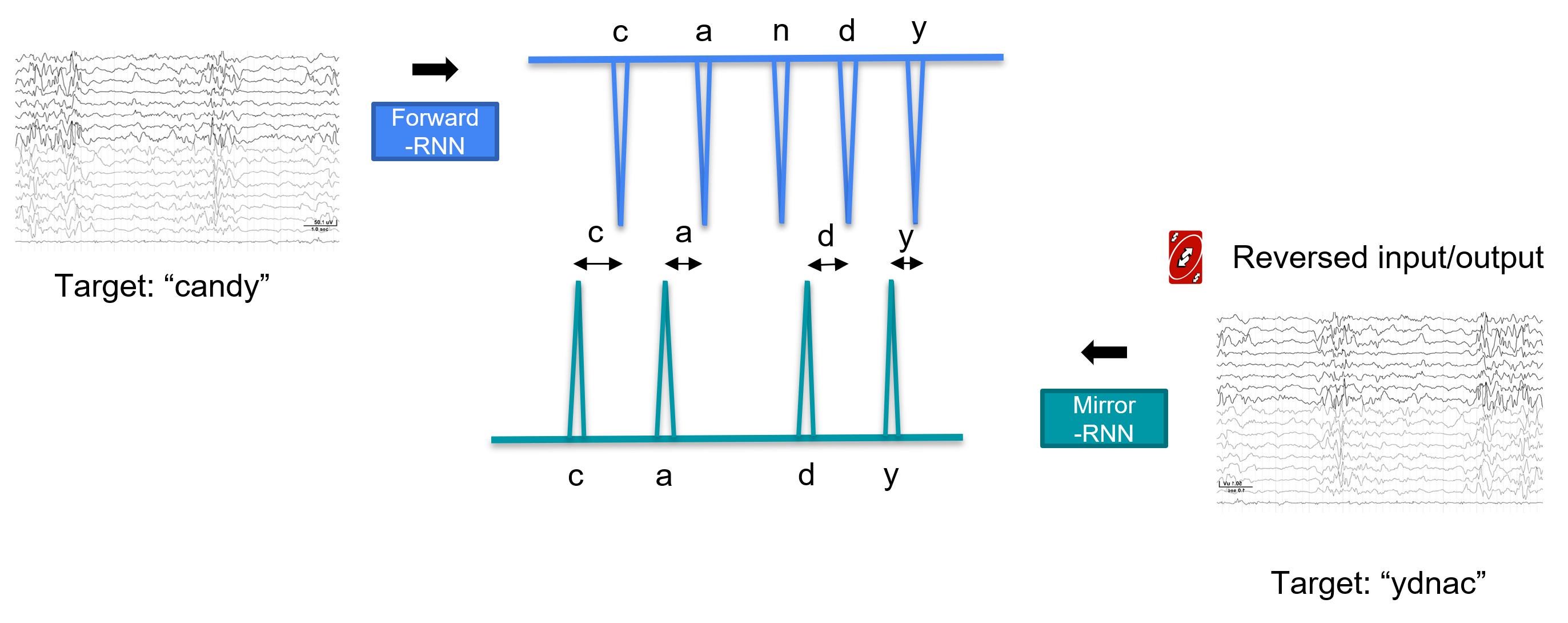}
    \caption{Method to extract character-level snippets from neural data. The forward-RNN is a regular character recognition RNN trained using CTC, and produces peaks when character ends are recognized. We train a second `Mirror-RNN' with reversed inputs and outputs. This model produces peaks at the start of characters, providing intervals for snippet extraction.}
    \label{fig:ctc_segmentation}
\end{figure*}

In this work, we introduce a simple method to use causal CTC models to achieve segmentation. 
Consider a continuous sentence writing example, in which the participant imagines writing the word `candy'. The example is illustrated in Fig \ref{fig:ctc_segmentation}. Neural data is usually fed to a recurrent model trained with CTC loss. In this example, we call it the forward model. As mentioned previously, the model produces a sharp probability peak for each target character that has been recognized. At all other times, the probability of the special `blank' character $\phi$ is highest. When a character peak is observed, we know only that the character was recognized due to some segment of input before this peak, due to the causal nature of the model. Thus, the forward model recognizes the ends of character segments, but does not indicate their beginnings.
% Typically, this sequence of recognized characters is taken as the output of the model, without precise alignment.

Next, we train a second recurrent model with the CTC loss. This model is architecturally identical to the forward model, but is trained on temporally reversed neural inputs as well as character targets.   
We refer to this model as the `mirror-RNN'. As shown in Figure \ref{fig:ctc_segmentation}, during inference time, we feed the neural data to the model reversed in time. Since the mirror-RNN is also a causal model, the probability peaks of the characters indicate the ending of \textit{reversed} character segments. We have thus identified the beginning of the characters in the input.

The forward and mirror models produce the ends and starts of each character respectively, and we extract neural snippets corresponding to each character using these intervals. Note that the models might produce different outputs since their inference is run unconstrained by each other (eg. in Fig \ref{fig:ctc_segmentation}, `n' is not produced by the mirror model). Thus, we obtain snippets only for the characters that both models produce, which are considered high-confidence outputs.

We can thus train and recalibrate our kinematics decoder without the need for any single-letter data collection. We now describe the architecture of the kinematics prediction model.

\begin{figure}[h]
    \centering
\includegraphics[width=0.4\textwidth]{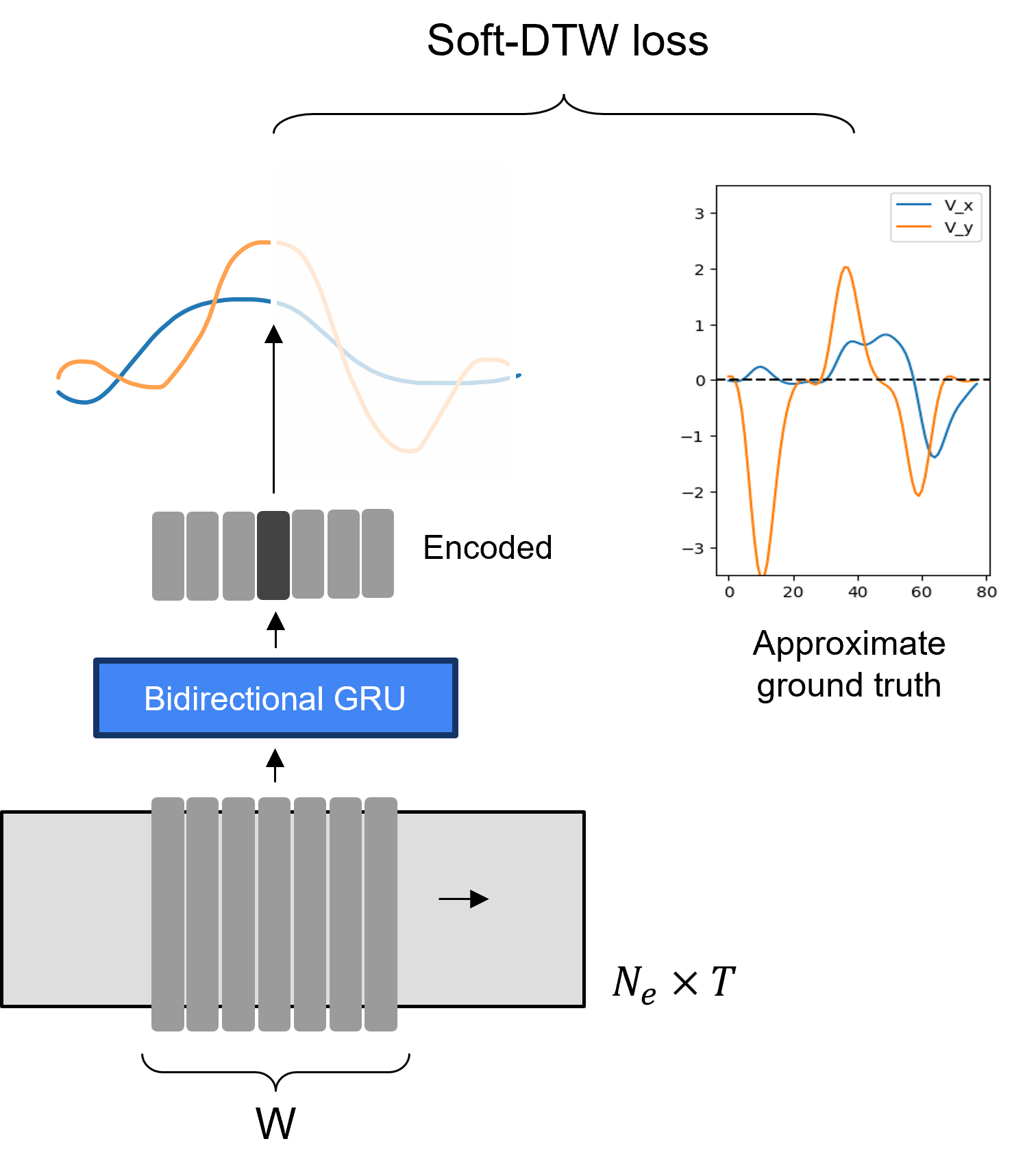}
    \caption{Kinematics prediction model architecture}
    \label{fig:kinematics_prediction}
\end{figure}

The kinematics prediction model and training is illustrated in Figure \ref{fig:kinematics_prediction}. Each example consists of a neural snippet of size $N_e \times T$, and the template trajectory of the character associated with the snippet. The model is a sequence model, which in this case is a Gated Recurrent Unit (GRU) layer introduced in \citet{cho2014properties}. The sequence model takes as input a window of neural data (of size $N_e \times W$). The output of the kinematics model is the instantaneous velocity (a $2d$ vector containing cartesian velocities $V_x$ and $V_y$) corresponding to the middle of the neural window. The window size W is intentionally chosen to be much smaller than the size of any letter's neural snippet seen during training. This is to ensure that the sequence model does not overfit to the trajectories observed during training, and can generalize in a zero-shot setting.
Upon sliding the window through the neural snippet, we obtain a predicted kinematic profile of size $2 \times T$.

In our experiments we found that symmetric context around the target resulted in best decoding performance. Thus we used a bidirectional GRU, at the cost of introducing a latency of $W/2$. Since $W$ is kept small, the effective latency introduced here is $\sim 80$ms for the results reported below. 

To demonstrate proof of concept, we extract all snippets of neural data corresponding to each letter from existing neural data collected in \citet{willett2021high} for training, and evaluate zero-shot recognition using a template matching approach. 

Many letters have distractors that have similar kinematic profiles, requiring fine-grained kinematics prediction. 
% We evaluated zero-shot recognition of `b' across multiple sessions using a template-matching approach.
Note that this is a hard version of the problem, since a realistic setting can use contextual information and a language model to improve performance.

\subsection{Evaluation}

% \textbf{Velocity Correlation}
% To quantify kinematic fidelity, we compute the Pearson Correlation Coefficient between the predicted and ground-truth 2D velocity traces. Let $\mathbf{V}, \hat{\mathbf{V}} \in \mathbb{R}^{2 \times T}$ denote the ground-truth and predicted velocity matrices. We calculate the correlation $\rho_x$ and $\rho_y$ for each dimension, and report the average correlation across dimensions: $\text{Score}_{\text{corr}} = \frac{1}{2}(\rho_x + \rho_y)$.

% \textbf{Template matching accuracy:} 

To measure kinematics decoding performance for any individual letter, we can inspect the soft-DTW loss between the predicted kinematics profile and the ground-truth template. This metric is independent of other trajectories in the language set.

To evaluate the final recognition performance for these characters, we compute a template matching accuracy. Given a dictionary of characters for which we have approximate ground-truth velocity templates $\{y_1, y_2 ... y_m\}$, and the predicted velocity profile $\hat{y}$ from our kinematics model, we can compute and rank each candidate using a DTW-based distance to the predicted velocity profile. 

\vspace*{-.4cm}
$$d_i = DTW(\hat{y}, y_i)$$
We compute hits@1 (accuracy) and hits@3 metrics based on whether the ground truth is in the top-1 or top-3 ranked candidates respectively. 

% \subsection{Grad-CAM analysis}

% While the RNN performed best for decoding, we wished to inspect the sources driving decoding performance for interpretability. We thus replaced the RNN decoder in the kinematics prediction model with a CNN, which can be more readily interpreted using existing techniques from the explainable AI field.
% Concretely, to interpret the spatiotemporal features prioritized by the CNN for kinematics decoding, we employed Gradient-weighted Class Activation Mapping (Grad-CAM). This technique produces a saliency map by utilizing the gradient information flowing through the network to assign importance values to specific neural features.

\subsection{Implementation}
\label{sec:implementation}

Both the forward and mirror classification models used an identical architecture. 
Following \citet{willett2021high, fan2023plug}, we used a single RNN common across all sessions, but employed day-specific linear projections to align cross-session neural data.
Both forward and mirror models were trained using the CTC loss. 
The kinematics model trained a day-specific single-layer RNN with 512 hidden units.
The input window length $W$ was chosen to be 7 for all reported decoding models, with time bins of 20ms. Soft-DTW loss was computed using $\gamma$ of $1e^{-5}$. 

All models were trained with a batch size of $64$, using the Adam optimizer and an L2 weight decay of $1e^{-5}$. The learning rate was set to a maximum of $1e^{-3}$, with a scheduler for linear warmup.

For results with clustering analyses, data was first projected using Principal Component Analysis (PCA) to $50$ dimensions, and t-distributed Stochastic Neighboring Embeddings (t-SNE) projections are computed using a perplexity of $30$.
All experiments were conducted on a single A6000 GPU. 
\section{Results}
\label{sec:results}

\subsection{Extracted neural snippets cluster along character}

\begin{figure}[h]
    \centering
\includegraphics[width=0.4\textwidth]{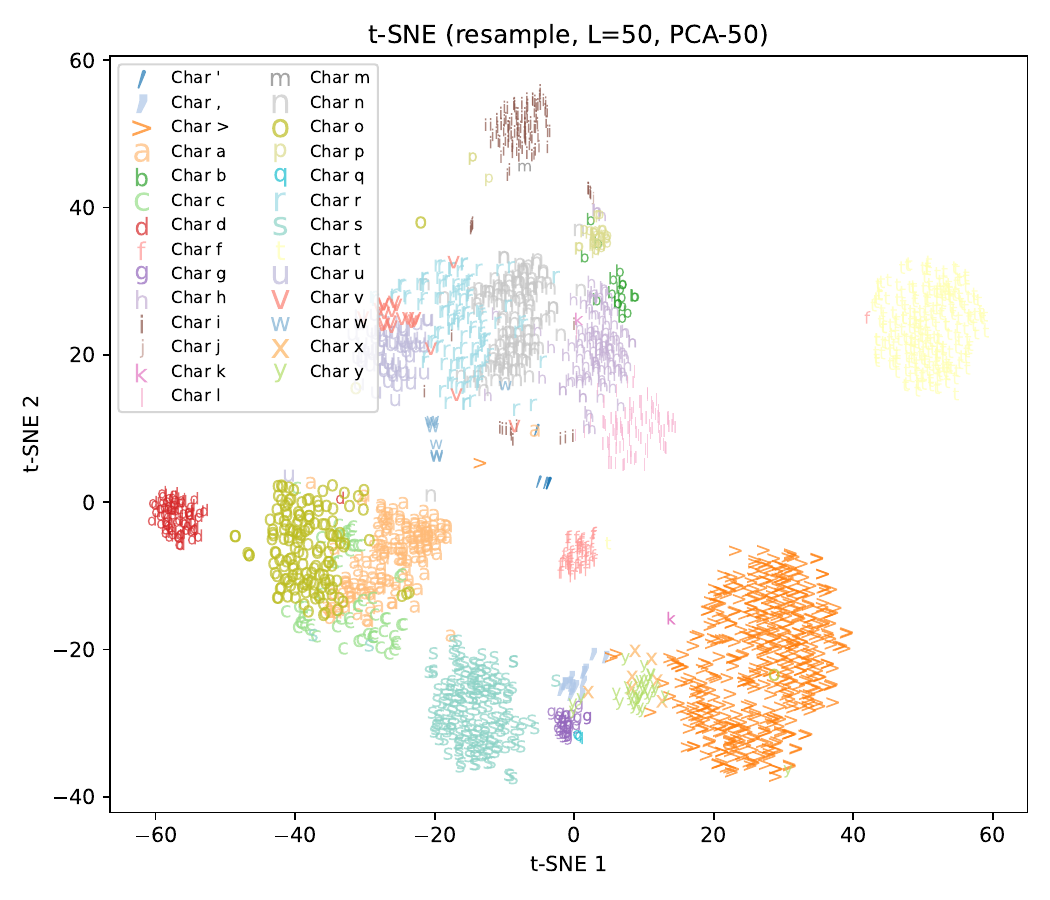}
    \caption{Extracted neural snippets in one session cluster  by character, validating our simple method to extract character snippets.}
    \label{fig:snippet_cluster}
\end{figure}

We first evaluate the snippets extracted by our novel method. As a sanity check, we inspect if the neural snippets associated with each character cluster together. Variable-sized snippets are resampled to the same length, and projected to 2 dimensions using PCA and t-SNE, as described in the previous section. Projected snippets are shown in Figure \ref{fig:snippet_cluster}, with each character denoted by a unique color. We observe that the extracted neural snippets clearly cluster by character. More accurately, the snippets cluster by kinematic profiles, with character snippets that have similar trajectories closer together.

\subsection{Predicted kinematics for zero-shot snippets can be correctly recognized}

\begin{figure}[h]
    \centering
\includegraphics[width=0.4\textwidth]{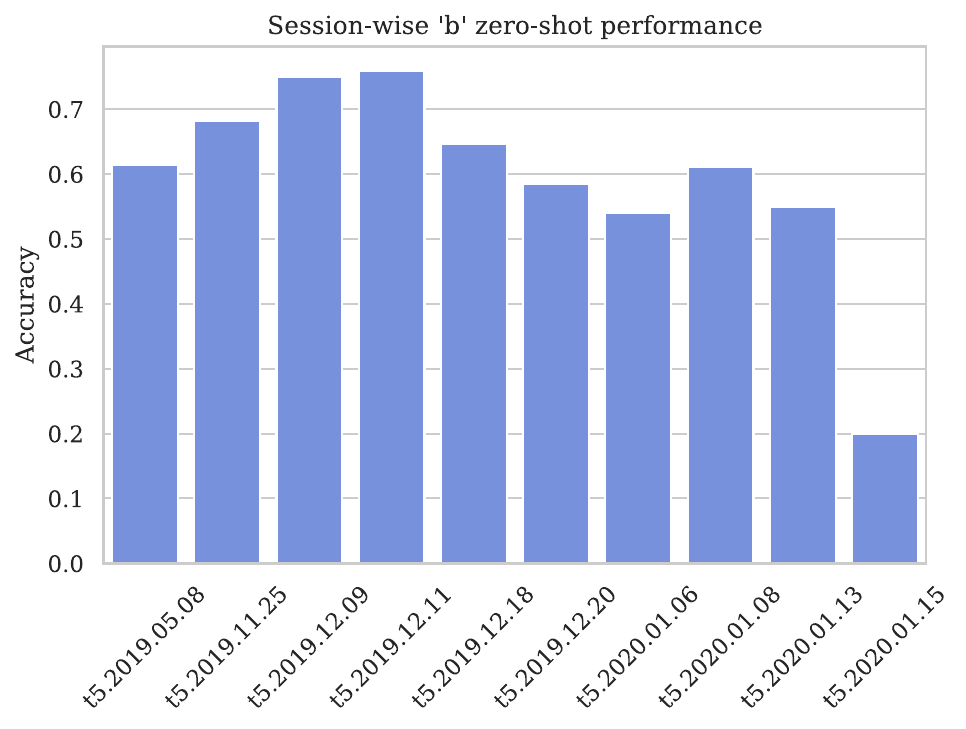}
    \caption{Session-wise zero-shot recognition of the letter `b'}
    \label{fig:b_zero_shot_perf}
\end{figure}

Next, we inspect whether the kinematics of a single held-out letter can be well predicted, as well as how this performance varies across sessions. We consider a random letter \texttt{`b'} for this analysis. 
As shown in Fig \ref{fig:b_zero_shot_perf}, we achieve up to 74\% recognition accuracy in the best session. 
This result demonstrates that our framework can achieve zero-shot recognition, with performance relatively stable across sessions. The performance drop in the last session might be attributed to the low number of trials (and consequently extracted snippets) in it for training the kinematics decoder. 

\begin{figure}[h]
    \centering
\includegraphics[width=0.4\textwidth]{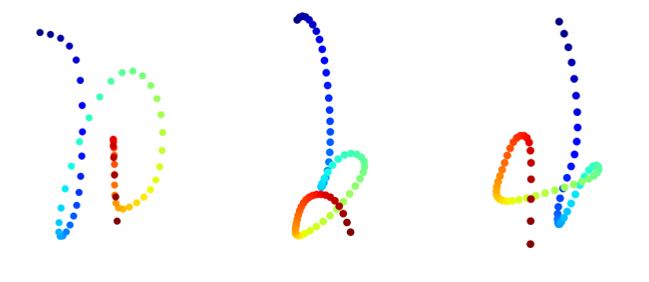}
    \caption{Three representative examples of predicted trajectories from single-trials of zero-shot letter `b', progressing from blue to red. Many of the predictions form human-recognizable trajectories.}
    \label{fig:b_recons}
\end{figure}

The velocity traces can integrated over time to visualize reconstructed trajectories. Some representative examples of single-trial predicted kinematics are shown in Fig \ref{fig:b_recons}, demonstrating highly recognizable shapes. This might allow human recognition of the kinematics even if the template matching stage fails.

\subsection{DTW optimal path shows consistent differences between template and real trajectories}

\begin{figure}[h]
    \centering
\includegraphics[width=0.4\textwidth]{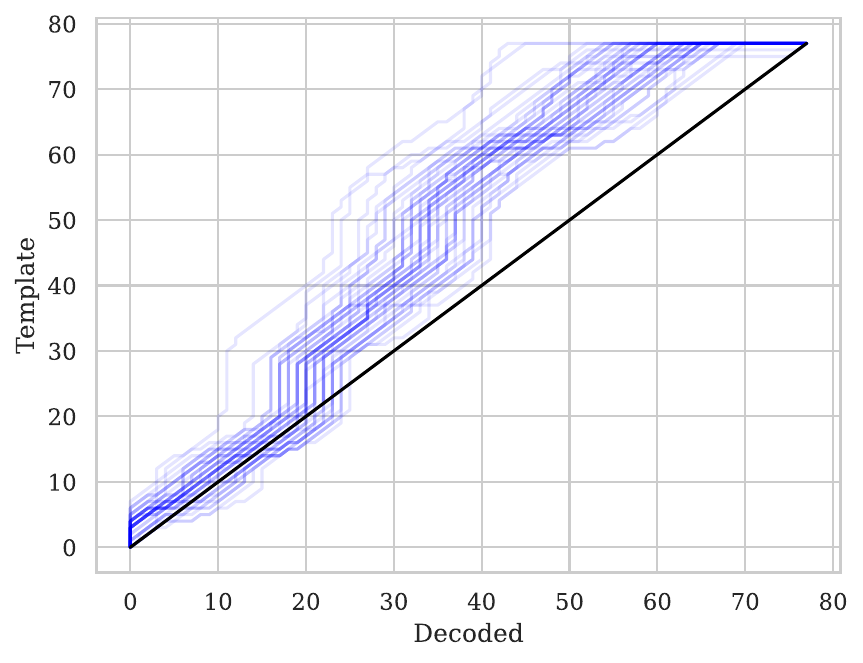}
    \caption{Optimal warping paths between predicted and template kinematics for various trials of the letter `b'. The predictions are internally consistent with one another, as well as consistently deviating from the diagonal.}
    \label{fig:b_warping_traces}
\end{figure}

With the DTW loss, we can visualize the optimal path that minimizes differences between the predicted kinematics and the velocity template.
These optimal paths are shown in blue in Figure \ref{fig:b_warping_traces}. Each path is read from the bottom left to the top right, and indicates at every timestep how much of the prediction and the template are `consumed' respectively in the matching process up to that timestep. When the optimal path moves diagonally, both the predicted kinematics and the template for that timestep are matched, indicating good alignment. If the optimal path moves vertically upwards, it indicates that multiple samples of the template are matched to the same point in the predicted velocity profile.

We observe that all paths (a) are consistent with each other, indicating that the participant imagines writing the letter with  stereotyped kinematics, and (b) consistently diverge from the diagonal, confirming that the approximate ground truth does not accurately capture the imagined velocity profile. This provides an opportunity for future work to update the ground truth template and repeat kinematics training, iterating until convergence.

\subsection{Character recognition performance varies across letters}

\begin{figure*}[t]
    \centering
    \includegraphics[width=\textwidth]{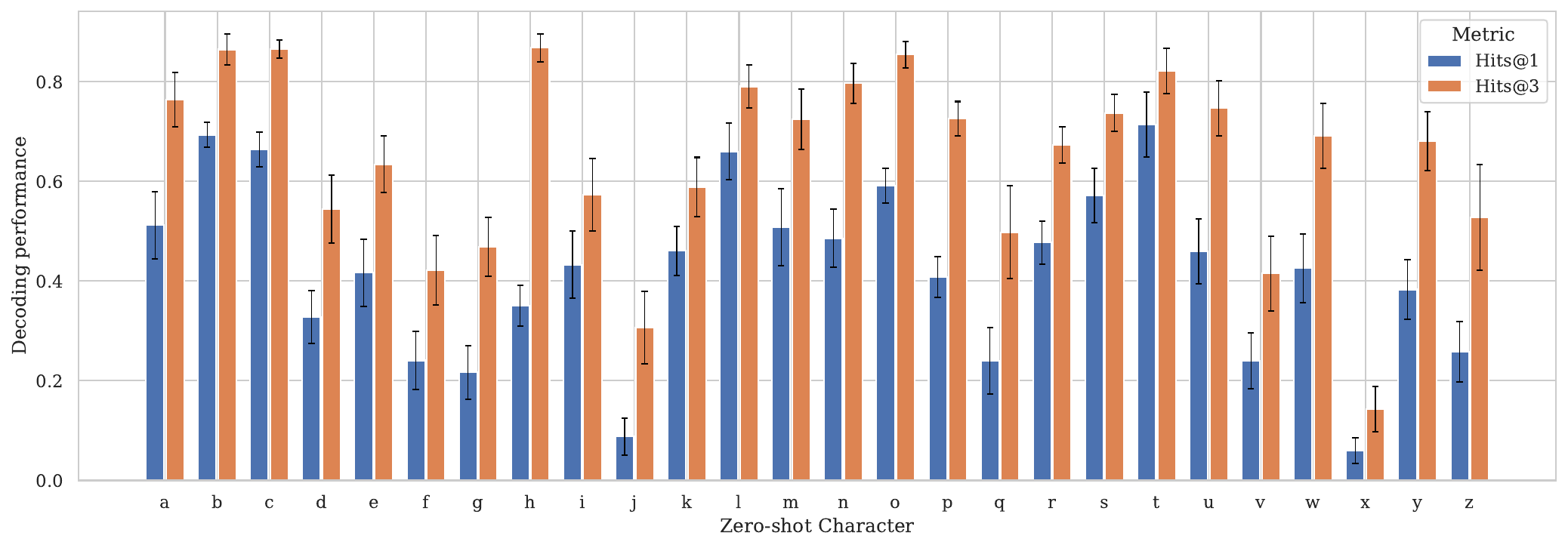}
    \caption{Mean decoding performance (hits@1 and hits@3) and standard error over sessions across held-out characters. We find that zero-shot decoding performance varies significantly between characters. 
    % \textcolor{red}{this is underestimating perf because some sessions computed as 0}
    }
    \label{fig:all_zero_shot_perf}
\end{figure*}

The zero-shot classification performance across all letters in the dataset is shown in Figure \ref{fig:all_zero_shot_perf}. We obtain an average of $41.88\%$ hits@1, and $64.35\%$ hits@3 across all held-out characters in this dataset. However, we observe significant differences in decoding performance across letter trajectories.

We identify two factors that may play a role in causing these large variations in performance.
First, the kinematic vectors of the held-out letter must be spanned by the kinematic vectors of all other letters, to effectively learn to predict the held-out trajectory. Some characters may have velocities that are not well-represented in the rest of the characters. This corresponds to a failure in the first stage (kinematics prediction).
Second, some letters might have harder distractors than others. This corresponds to failures in the second stage (template matching).

\begin{figure}[h]
    \centering
\includegraphics[width=0.45\textwidth]{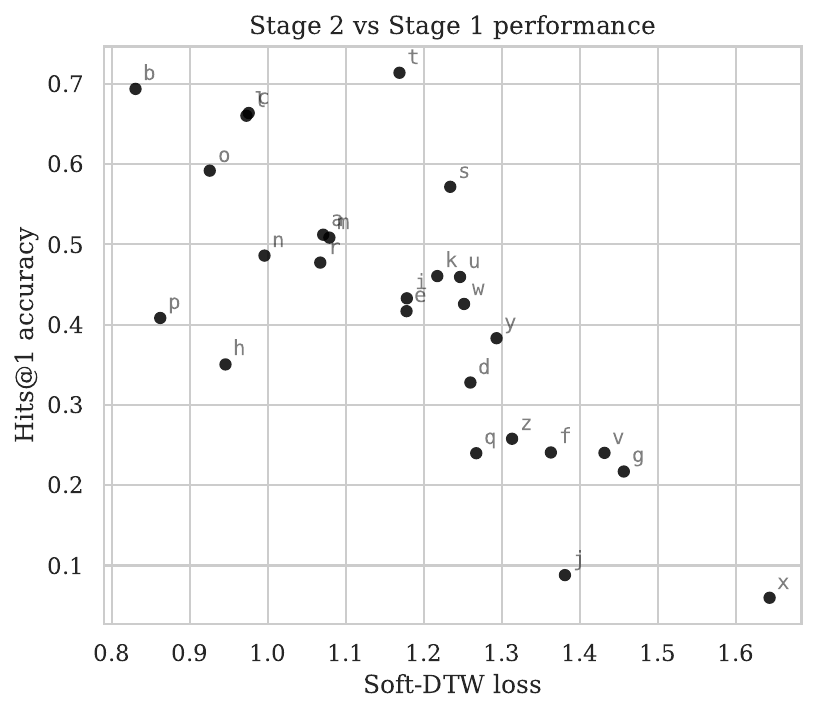}
    \caption{Comparing character-wise performance after full pipeline vs performance of Stage 1 (Kinematics prediction) alone, as measured by Hits@1 and Soft-DTW loss respectively. In the bottom-right quadrant most letters appear limited by Stage 1 performance (kinematics prediction), while errors in Stage 2 (template matching) contribute to the lower-than-expected performance for certain letters in the left half (eg. p, h).
    }
    \label{fig:softDTW_loss}
\end{figure}

We inspect a scatter plot of performance post-Stage 1 (Soft-DTW loss) and post-Stage 2 (Hits@1). A strong correlation would suggest that variability in performance is driven by Stage 1 failures, while weak correlation would suggest Stage 2 failures. As shown in Figure \ref{fig:softDTW_loss}, we discover an interesting result : both sources of error contribute to this variability, but at different regimes of Hits@1. We observe a tighter correlation between final performance and Stage 1 performance in the worst performing characters. However, certain letters like ‘p’ and ‘h’ are very well decoded in Stage 1, but are likely impacted by difficult distractors in Stage 2.

% \begin{figure}[h]
%     \centering
% \includegraphics[width=0.4\textwidth]{Figures/SoftDTW_losses.pdf}
%     \caption{Kinematic decoding performance across characters as measured by the Soft-DTW loss to the original template indicate comparable velocity prediction performance. }
%     \label{fig:softDTW_loss}
% \end{figure}

\begin{figure}[h]
    \centering
\includegraphics[width=0.4\textwidth]{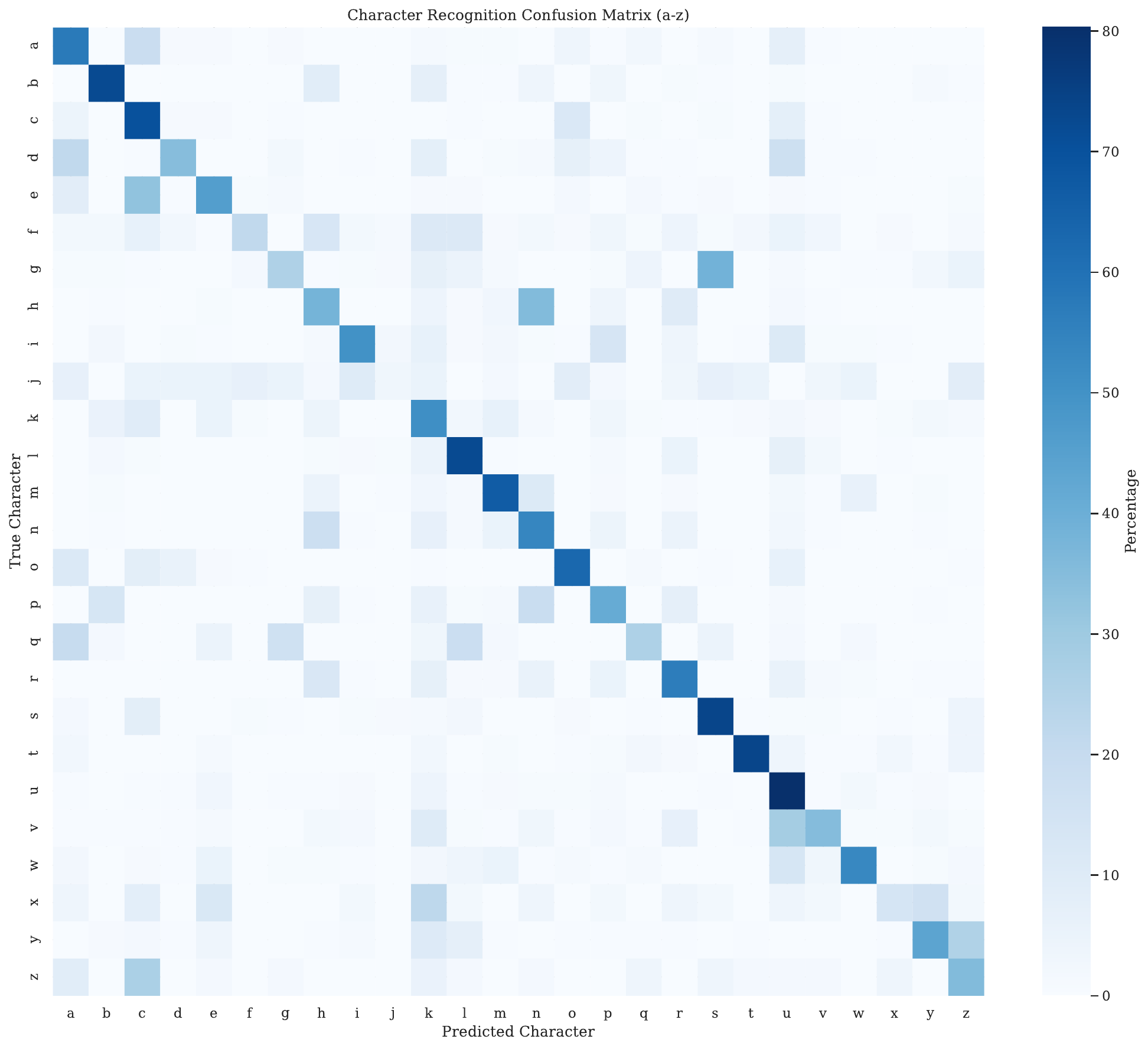}
    \caption{Normalized confusion matrix across held-out characters with different motor patterns, which confirms that errors occur primarily in pairs of letters with similar velocity profiles.}
    \label{fig:confusion_matrix}
\end{figure}

To investigate further, we inspect the normalized confusion matrix for zero-shot character recognition, shown in Figure \ref{fig:confusion_matrix}. We observe that character pairs with similar kinematic trajectories are indeed confused more often, such as \texttt{(`a', `c')},  \texttt{(`e', `c')}, \texttt{(`h', `n')}, and \texttt{(`g', `s')}. While \texttt{`j'} achieved the lowest decoding accuracy and appears to be uniformly confused with all characters, we find that a very limited number of snippets (3 to 13 across sessions) were extracted, potentially leading to high-variance evaluation. 

% Both factors will vary with language. 
% However, it is notable that a straightforward indicator of kinematic decoding performance, the soft-DTW loss to the original velocity template of the character, is comparable across letters, as shown in Figure \ref{fig:softDTW_loss}. 
% In particular, we highlight how the kinematic predictions of \texttt{`j'} or \texttt{`x'} are quite high, independent of their final classification performance due to possible distractors.  This provides further evidence that some neural representation of kinematics is robustly conserved across character contexts, since the kinematics can be uniformly well predicted. 

\subsection{Kinematics prediction exhibits lower cross-session stability than character classification}

\begin{figure}[h!]
    \centering
\includegraphics[width=0.4\textwidth]{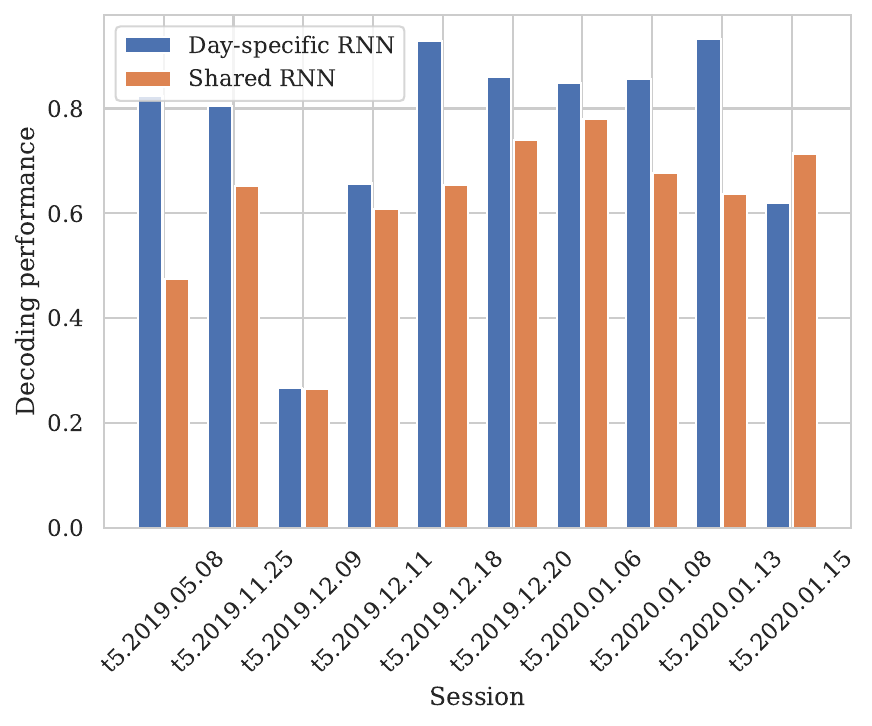}
    \caption{Day-specific kinematics decoders (RNNs) trained separately on each day's neural data outperform a shared RNN with day-specific linear projections, which suffices for character classification models, suggesting more representational drift for low-level kinematics.}
    \vspace{-3mm}
    \label{fig:kinematics_representational_drift}
\end{figure}

We observed a significant divergence in cross-session generalization between the continuous kinematics decoding task and the CTC-based models that perform discrete character classification from prior work. For character classification, aligning neural activity via day-specific linear projections was sufficient to stabilize performance, enabling a single, shared RNN decoder to operate effectively across all sessions \citep{willett2021high, fan2023plug}. In contrast, this alignment strategy proved inadequate for kinematics prediction. As shown in Figure \ref{fig:kinematics_representational_drift} on an example letter, training separate, day-specific RNNs outperformed a shared RNN. 

This discrepancy suggests that the neural representation of continuous motor kinematics exhibits greater inter-session variability than  features governing discrete character identity. While the decision boundaries for classification appear to be linearly alignable across days, the precise dynamic relationship between neural activity and hand velocity likely undergoes complex, non-linear drift, requiring  day-specific model adaptation.

% \subsection{CNN decoder interpretation}

% The Grad-CAM results are shown in Figure.

% \subsection{Continuous decoding performance}

% Finally, to achieve high throughput communication as in the original work \cite{willett2021high}, zero-shot recognition must be achieved in a continuous setting. 

% First, we confirm that the 
\section{Discussion}
\label{sec:discussion}

This work is meant to highlight the need for zero-shot character recognition capability in handwriting BCIs, and investigate whether the neural basis of handwriting can facilitate the same. We provide a proof-of-concept demonstration that a shared kinematics representation is robustly conserved across diverse character contexts and can be leveraged to achieve zero-shot recognition.

To do so, we introduce a novel method to 
train and automatically recalibrate a kinematics model without the need for supervised single-letter data collection. We use the CTC-trained model that a subject would employ for their everyday BCI use, to extract snippets of neural data associated with each character from any given day's neural data. Our framework thus facilitates the training of a kinematics decoder that can adapt to imperfect ground truth kinematic profiles.

% While HMMs can and have been used \cite{willett2021high} to segment the neural data, CTC-trained RNN models have been shown to be more robust to distribution shifts across sessions. Moreover, prior work using HMMs to segment neural data into character snippets used supervised single-letter data and manual parameter tuning. 

We validated our snippet extraction procedure by demonstrating that neural snippets cluster by motor trajectory.
Our framework provides DTW-based optimal frame-wise alignment between kinematics and neural recordings in large datasets, without the need for a slow supervised experimental paradigm.

We reported $41.88\%$ hits@1 and $64.35\%$ hits@3 mean recognition performance across all held-out characters in the chosen dataset, demonstrating conserved kinematics representations across characters. Upon visualization, decoded character trajectories were human recognizable in many cases.

Interestingly, we observed divergence in cross-session stability between continuous kinematics prediction and discrete character classification. This highlights a distinction in how different aspects of motor behavior are maintained in the cortex over time. Our findings may be explained by two competing hypotheses:

(a) While the neural manifolds governing abstract, categorical features such as character identity are sufficiently stable to be aligned via simple linear transformations, the fine-grained representations of continuous hand velocity are subject to more complex, non-linear representational drift. This discrepancy could reflect the hierarchical organization of motor control \citep{merel2019hierarchical}. High-level movement intentions and discrete motor programs appear robustly preserved and linearly accessible across days. In contrast, it is possible that lower-level dynamic execution signals are continuously adapting to fluctuations in background network states or even ongoing motor consolidation, rendering simple linear alignment strategies inadequate. 

% (b) Alternatively, this divergence may arise from the physical realities of the recording interface. High-level features are often widely distributed and highly redundant, making them robust to minor signal loss. In contrast, the fine-grained decoding of continuous kinematics relies on specific, localized populations, making it acutely sensitive to physical perturbations at the electrode-tissue interface, such as array micro-shifts or the localized loss of individual neurons over time.
(b) An alternative hypothesis is that this divergence is driven less by functional adaptation and more by the physical limitations of the recording interface. High-level categorical information is typically represented by widespread, redundant population activity, allowing it to be reliably decoded even if individual channels drop out. However, precisely decoding low-level kinematics depends on the exact firing rates of specific local neurons. Therefore, continuous decoding is far more susceptible to minor hardware disruptions, including micro-movements of the implanted array, local gliosis, or neuronal death at the electrode interface, all of which might manifest as non-linear representational drift.

In either case, from a translational perspective, the method described in this work can specifically address these complex, non-linear representational shifts. By facilitating the recalibration of the kinematics decoder as the participant uses their character recognition BCI, we can accommodate the day-to-day volatility of continuous motor representations. Our work demonstrates that while continuous trajectory decoding is inherently more susceptible to inter-session drift than discrete classification, unsupervised daily recalibration strategies can effectively provide a pathway toward robust, long-term continuous BCI control.

Our work is closely related to recent efforts \citep{chen2026surrogate, qi2025human} that employ state-dependent linear decoding of imagined handwriting. \citet{qi2025human}  explicitly identifies states that affect tuning properties of individual neurons in the MC, in order to learn a unique decoder per state. However, such approaches require manual estimation of the number of stable states, which may itself change over sessions. Our method is a generalization of such approaches, learning non-linear recurrent mappings directly in a data-driven manner.

% Finally, from the perspective of building a functional handwriting BCI, it should be noted that a language model can likely solve the zero-shot recognition problem using contextual cues alone for a small set of unseen letters. We emphasize that we do not seek to replace language models post-processing the output to produce corrections. Rather, when the number of zero-shot characters becomes large, less context can be decoded with a vanilla approach, and a zero-shot capable model is useful before a language model is applied.

A limitation of this work is that it still requires approximate ground-truth kinematics of a set of letters. Acquiring such approximate ground-truth can be challenging when the participant has already experienced significant paralysis. However, one approach to address this challenge could exploit letters (such as \texttt{`c'}) with limited variability in writing styles across the population. A few such letters could be used to bootstrap a kinematics decoder, using it to infer approximate ground truth for other letters. Future work will address this direction.

Finally, we emphasize the need for principled datasets collected specifically to evaluate zero-shot symbol recognition in handwriting decoding.
% and phoneme recognition in speech decoding. 
Although our offline demonstration serves as a proof of concept, the development of dedicated datasets will be critical for driving new methods in zero-shot generalization. 
% For instance, a standardized zero-shot evaluation paradigm could be established by training models exclusively on lowercase letters and evaluating them on a strictly held-out set of capital letters. 
Applying our framework to a non-character dataset presents a compelling direction for future research. Such a dataset would provide a unique opportunity to investigate whether the neural representations of continuous kinematics vary depending on the presence or absence of underlying semantic meaning.

\printbibliography

\end{document}